\documentclass[aps,prl,twocolumn,superscriptaddress,showpacs]{revtex4-1}
\usepackage{natbib}
\usepackage{amsmath}

\usepackage{color}
\usepackage{graphicx}
\usepackage{dcolumn}
\usepackage{bm}
\usepackage[normalem]{ulem}
\usepackage{multirow}

\begin{document}

\preprint{APS/123-QED}

\title{Giant Anomalous Hall Effect in the Chiral Antiferromagnet Mn$_3$Ge}

\author{Naoki Kiyohara}
\author{Takahiro Tomita}
\affiliation{The Institute for Solid State Physics, The University of Tokyo, Kashiwa, Chiba 277-8581, Japan}
\author{Satoru Nakatsuji}
\email{satoru@issp.u-tokyo.ac.jp}
\affiliation{The Institute for Solid State Physics, The University of Tokyo, Kashiwa, Chiba 277-8581, Japan}
\affiliation{PRESTO, Japan Science and Technology Agency (JST), 4-1-8 Honcho Kawaguchi, Saitama 332-0012, Japan}
\affiliation{CREST, Japan Science and Technology Agency (JST), 4-1-8 Honcho Kawaguchi, Saitama 332-0012, Japan}

\date{\today}

\begin{abstract}
The external field control of antiferromagnetism is a significant subject both for basic science and technological applications. As a useful macroscopic response to detect magnetic states, the anomalous Hall effect (AHE) is known for ferromagnets, but it has never been observed in antiferromagnets until the recent discovery in Mn$_3$Sn. Here we report another example of the AHE in a related antiferromagnet, namely, in the hexagonal chiral antiferromagnet Mn$_3$Ge. Our single-crystal study reveals that Mn$_3$Ge exhibits a giant anomalous Hall conductivity $|\sigma_{xz}| \sim~60~\Omega^{-1}$ cm$^{-1}$ at room temperature and approximately $380~\Omega^{-1}$ cm$^{-1}$ at 5 K in zero field, reaching nearly half of the value expected for the quantum Hall effect per atomic layer with Chern number of unity. Our detailed analyses on the anisotropic Hall conductivity indicate that in comparison with the in-plane-field components $|\sigma_{xz}|$ and $|\sigma_{zy}|$, which are very large and nearly comparable in size, we find $|\sigma_{yx}|$ obtained in the field along the $c$ axis is found to be much smaller. The anomalous Hall effect shows a sign reversal with the rotation of a small magnetic field less than 0.1 T. The soft response of the AHE to magnetic field should be useful for applications, for example, to develop switching and memory devices based on antiferromagnets.
\end{abstract}

\maketitle

\section*{I. Introduction}
The anomalous Hall effect (AHE) is one of the best-studied transport properties of solid. Since its discovery in 1880, the effect is known to be proportional to magnetization, and, thus, the zero field AHE has been observed only in ferromagnets \cite{chien,nagaosa2010}. Hypothetically, however, since intrinsic AHE arises owing to fictitious fields due to Berry curvature, it may appear in spin liquids and antiferromagnets without spin magnetization in certain conditions, even with a large Hall conductivity comparable with the quantum Hall effect (QHE) \cite{Shindou2001,Bruno2006,Martin2008,Yang2011,Ishizuka2013,Chen2014,Kubler2014}.  Indeed, a spontaneous Hall effect has been observed in recent experiments in the spin liquid Pr$_2$Ir$_2$O$_7$ \cite{Machida2010} and the antiferromagnet Mn$_3$Sn \cite{Mn3Sn}. Nonetheless, the zero-field AHE observed to date reached only a few orders of magnitude lower value than the QHE per atomic layer. 

In recent years, antiferromagnets have attracted an increasing amount of attention due to the useful properties, in particular, for spintronics \cite{MacDonald2006, Jungwirth2010,MacDonald2011,Park2011,Marti2014,Gomonay2014}. In contrast with ferromagnets that have been mainly used to date \cite{Fert2007}, antiferromagnets are much more insensitive against magnetic field perturbations, providing stability for the data retention. In addition, antiferromagnets produce almost no stray fields that perturb the neighboring cells, removing an obstacle for high-density memory integration. Moreover, antiferromagnets have much faster spin dynamics than ferromagnets, opening avenues for ultrafast data processing. 

On the other hand, to develop antiferromagnetic devices, it is necessary to find detectable macroscopic effects that can be changed by the rotation of the sublattice moments. Thus, if we can find an antiferromagnet that exhibits a large AHE at room temperature, it will be useful for switching and memory devices, as a large change in the Hall voltage clearly defines binary information.

In this article, we report the observation of a giant anomalous Hall conductivity in an antiferromagnet reaching approximately $50$\% of the layered quantum Hall effect with Chern number of unity. In particular, we show that the noncollinear antiferromagnet Mn$_3$Ge isostructural to Mn$_3$Sn exhibits strikingly large anomalous Hall conductivity in zero field of approximately 60 $\Omega^{-1}$ cm$^{-1}$ at room temperature and approximately 380 $\Omega^{-1}$ cm$^{-1}$ at 5 K. Moreover, the sign of the giant AHE can be softly flipped by the rotation of magnetic field, indicating that the direction of a fictitious field equivalent to more than $200$ T is tunable by a small external magnetic field less than 0.1 T. Thus, the AHE should be useful for applications, for example, to develop switching and memory devices based on antiferromagnets.

\begin{figure}[t]
	\begin{center}
		\includegraphics[width=\columnwidth]{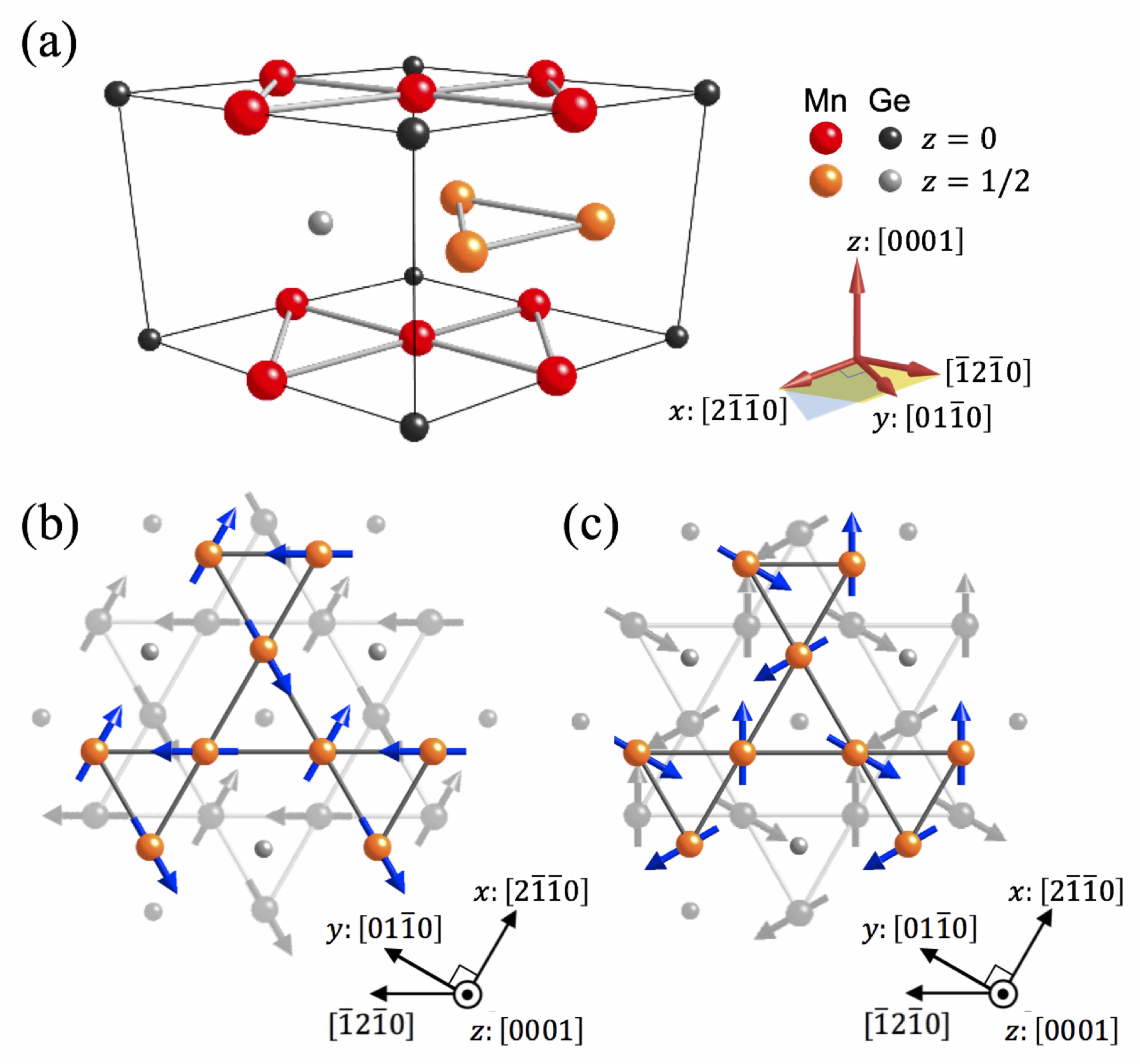}
		\caption{ {Crystal and magnetic structures of Mn$_3$Ge.} (a) Unit-cell crystal structure. To distinguish Mn and Ge on different $x$-$y$ planes with  $z = 0, 1/2$, those atoms in different planes are shown by different colors. (b),(c) Mn atoms form a breathing-type Kagome lattice, and their spins have a 120$^\circ$ magnetic structure as shown by blue arrows. (b),(c) show the most likely magnetic structures on the neighboring two layers with $z = 0$ (gray), $1/2$ (color) in the field along $[2\bar{1}\bar{1}0]$ and $[01\bar{1}0]$, respectively \cite{Nagamiya1982,Tomiyoshi1983}. Here, we define $[2\bar{1}\bar{1}0]$, $[01\bar{1}0]$ and $[0001]$ as the $x$, $y$, and $z$ axes, respectively.} 
		\label{fig1}
	\end{center}
\end{figure}

\begin{figure*}[t]
	\begin{center}
		\includegraphics[width=17cm]{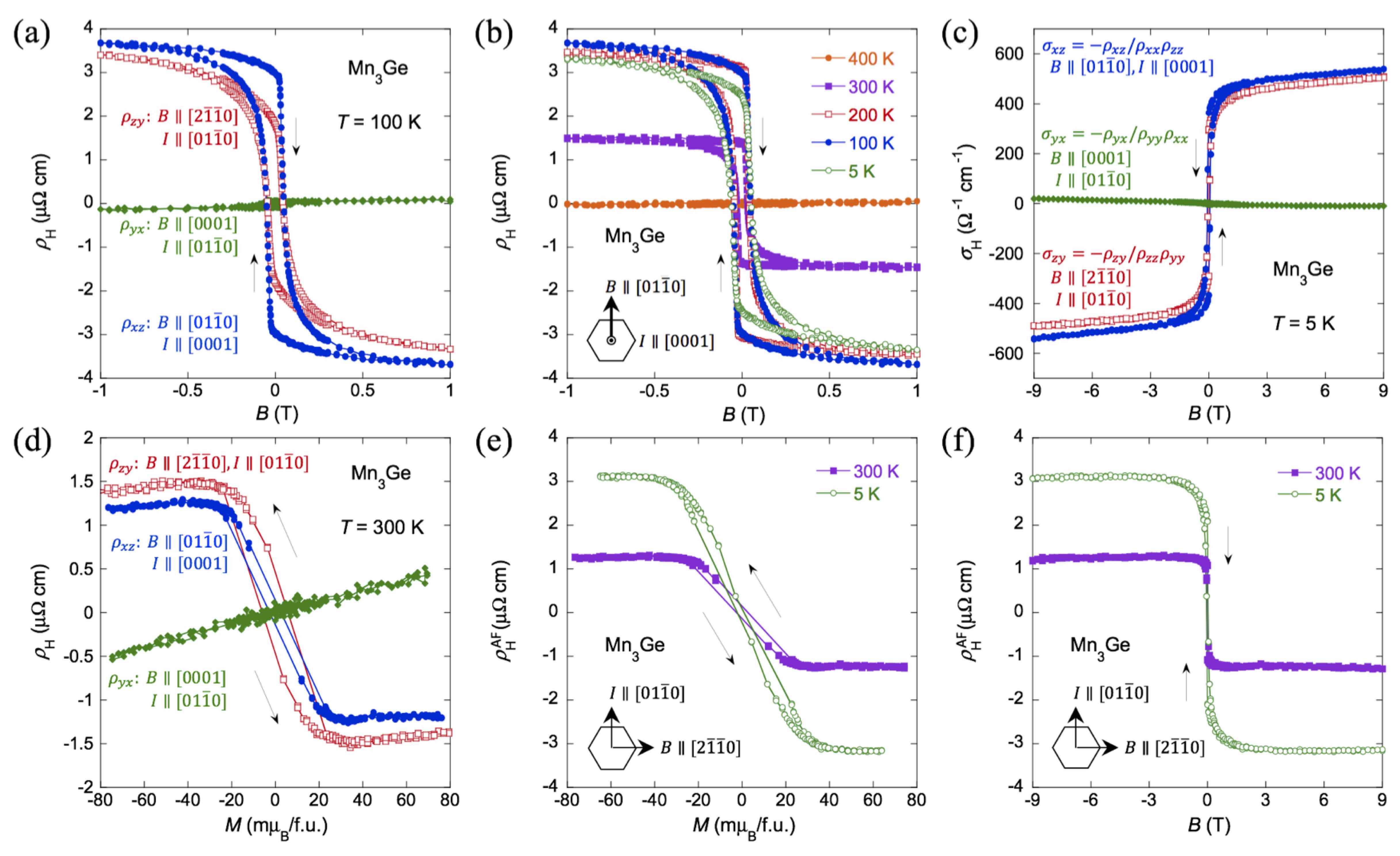}
		\caption{ Magnetic field and magnetization dependence of the anomalous Hall effect in Mn$_3$Ge. (a) \rm Magnetic field dependence of the Hall resistivity $\rho_\mathrm{H}$ measured in $B\parallel [2\bar{1}\bar{1}0]$, $[01\bar{1}0]$ and $[0001]$ at 100 K. (b) Magnetic field dependence of the Hall resistivity $\rho_\mathrm{H} =\rho_{xz}$ at 5, 100, 200, 300, and 400 K in $B\parallel [01\bar{1}0]$ with $I\parallel [0001]$. The hexagon and arrows at the lower left, respectively, show the hexagonal lattice and the field and current directions. (c) Magnetic field dependence of the Hall conductivity $\sigma_\mathrm{H}$. Directions of the field $B$ and electric current $I$ used for the Hall resistivity measurements are shown. (d) Magnetization dependence of $\rho_\mathrm{H}$ at 300 K measured in  $B\parallel [2\bar{1}\bar{1}0]$, $[01\bar{1}0]$, and $[0001]$. (e) Magnetization dependence of $\rho_\mathrm{H}^\mathrm{AF}=\rho_\mathrm{H}-R_0 B - R_s \mu_0 M$ at 5 and 300 K. (f) Magnetic field dependence of $\rho_\mathrm{H}^\mathrm{AF}$ at 5 and 300 K.} 
		\label{fig2}
	\end{center}
\end{figure*}

Mn$_3$Ge is isostructural to Mn$_3$Sn, which has Ni$_3$Sn-type structure with the hexagonal symmetry $P6_3/mmc$ [Fig. \ref{fig1}(a)]. The structure is stable only when there is excess Mn randomly occupying the Ge site. As a result, this phase exists over the range of Mn$_{3.2}$Ge-Mn$_{3.4}$Ge \cite{Yamada1988}. The projection of the Mn atoms onto the basal plane is a triangular lattice made by a twisted triangular tube of face-sharing octahedra. In each plane, the Mn atoms form a ``breathing'' type of a Kagome lattice (an alternating array of small and large triangles), and the associated geometrical frustration leads to a noncollinear 120$^\circ$ spin ordering of the magnetic moments approximately $3\ \mu_{\rm B}$/Mn below the N\'{e}el temperature of approximately 380 K, similarly to Mn$_3$Sn \cite{Nagamiya1982,Tomiyoshi1982}. Contrary to the usual 120 degree order, all Mn moments lying in the $x$-$y$ plane form a chiral spin texture with an opposite vector chirality owing to the Dzyaloshinskii-Moriya interaction [Figs. \ref{fig1}(b) and \ref{fig1}(c)]. This inverse triangular structure has the orthorhombic symmetry and induces an in-plane weak ferromagnetic (FM) moment of the order of approximately $0.007  \mu_{\rm B}/$Mn, which is believed to arise from the spin canting toward the local easy axis along the $[2\bar{1}\bar{1}0]$  direction \cite{Nagamiya1982,Tomiyoshi1983}. This in-plane chiral magnetic phase is stable down to the lowest $T$'s \cite{Yamada1988}, which allows us to observe a giant AHE at low temperatures, as we will discuss. In contrast, Mn$_3$Sn has a low-$T$ noncoplanar magnetic phase at $T < 50$ K, where the in-plane AHE is strongly suppressed \cite{Mn3Sn}. In our study, single crystals with the composition of Mn$_{3.05}$Ge$_{0.95}$ (Mn$_{3.22}$Ge) are mainly used and referred to as ``Mn$_3$Ge" for clarity (see Appendix A).

\section*{II. Results and Discussion}
We first present our main experimental evidence for the giant anomalous Hall effect found in Mn$_3$Ge. The Hall voltage is measured in the direction perpendicular to both the magnetic field $B$ and the electric current $I$. Figure \ref{fig2}(a) shows the field dependence of the Hall resistivity $\rho _{\rm H}$ obtained at 100 K in $B \parallel$ $[2\bar{1}\bar{1}0]$, $[01\bar{1}0]$, and $[0001]$ with $I$ perpendicular to $B$. It exhibits a clear hysteresis loop with a large change $\Delta \rho_{\rm H}\sim 5\ \mu\Omega$cm for $B \parallel [01\bar{1}0]$ comparable to Mn$_3$Sn \cite{Mn3Sn}. Besides, for free-electron gas with the carrier number estimated from $R_0$ (see below), it requires $B > 200$ T for the ordinary Hall effect to reach the observed values of $\Delta \rho_{\rm H}$ (see Appendix B). The hysteresis takes a similar small magnetic field to the Mn$_3$Sn case; the coercivity increases from 300 Oe at 300 K to 600 Oe at 5 K [Fig. \ref{fig2}(b)], while it remains constant at approximately $300$ Oe for Mn$_3$Sn \cite{Mn3Sn}. This large anomaly is seen only in $\rho _{\rm H}$. The magnetoresistance (ratio) in this $T$ range (Appendix C) is less than 0.6 $\mu \Omega$cm (0.4\%), which is 1 order of magnitude smaller than $\Delta \rho_{\rm H}$. We further find that the hysteresis in $\rho _{\rm H}$ is robust against a small change in the Mn concentration (see Appendix D). 

To clarify the mechanism of transport properties in general, it is important to find the associated anisotropy. In the study of the anomalous Hall effect, however, the anisotropy is largely neglected. Since the longitudinal resistivity is anisotropic at low $T$'s (see below), for the estimate of the Hall conductivity, we employ the expression $\sigma_{ji} \approx -\rho_{ji}/(\rho_{jj}\rho_{ii})$, where $(i,j)=(x,y),\  (y,z),\ (z,x)$ (see Appendix E). The results show a sharp hysteresis and reach large values of approximately $500~\Omega^{-1}\mathrm{cm}^{-1}$ at $T = 5$ K and $B = 9$ T [Fig. \ref{fig2}(c)]. This is nearly 4 times larger than in Mn$_3$Sn \cite{Mn3Sn}, and reaches approximately $60$\% of the value (approximately $800~\Omega^{-1}\mathrm{cm}^{-1}$) expected for a layered QHE as we discuss.  In contrast, both $\rho_{yx}$ and $\sigma_{yx}$ for $B \parallel [0001]$ exhibit a linear increase with $B$ except a very small hysteresis with $\Delta \rho_{\rm H} < 0.1~\mu \Omega$cm and $\Delta \sigma_{\rm H} < 4~\Omega^{-1}\mathrm{cm}^{-1}$ found around $B = 0$ [Figs. \ref{fig2}(a) and \ref{fig2}(c)]. Significantly, similar sharp and anisotropic change as a function of field is seen at 300 K, as shown in Appendix F.

\begin{figure}[t]
	\begin{center}
		\includegraphics[width=6.9cm]{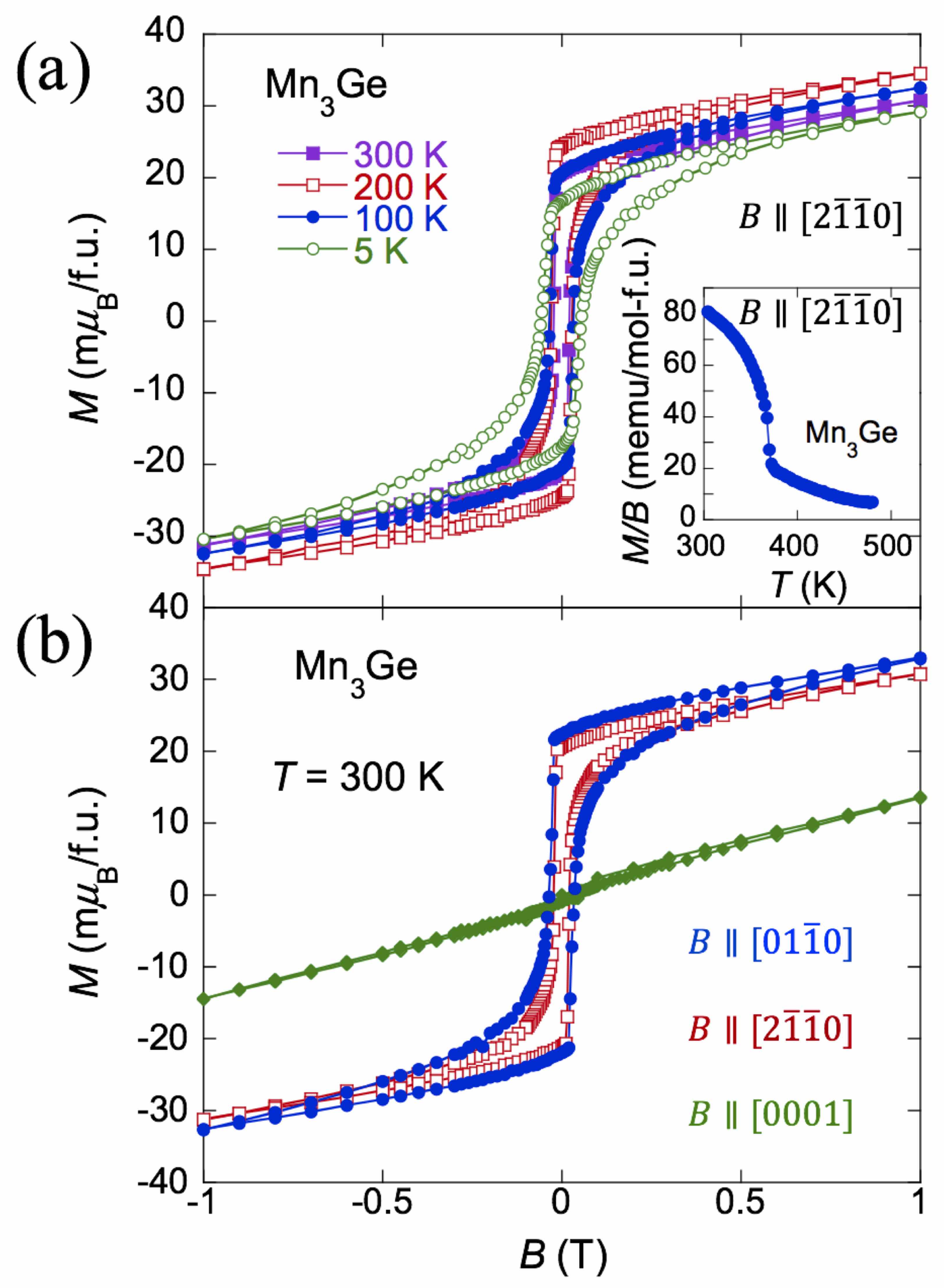}
		\caption{(a) Magnetic field dependence of the magnetization $M$ measured in $B\parallel[2\bar{1}\bar{1}0]$ at various temperatures. The inset indicates the temperature dependence of the susceptibility $M/B$ obtained above 300 K in $B = 0.1$ T $\parallel[2\bar{1}\bar{1}0]$ in the field-cooling procedure. (b) Magnetization curve obtained at 300 K in $B\parallel[2\bar{1}\bar{1}0]$, $[01\bar{1}0]$ and $[0001]$.} 
		\label{fig3}
	\end{center}
\end{figure}

This sign change with a large jump of the anomalous Hall conductivity most likely indicates that the direction of the sublattice moments switch in response to the change in the external field by approximately 0.1 T, suggesting an extremely small energy scale associated with magnetocrystalline anisotropy \cite{Nagamiya1982,Tomiyoshi1983}. Various spin configurations in the in-plane fields are shown in Appendix G. Indeed, a theoretical analysis reveals that the inverse triangular spin structure should have no in-plane anisotropy energy up to the fourth-order term \cite{Nagamiya1982,Tomiyoshi1983}. Thus, the spin triangle should rotate easily, following the sign change of magnetic field. Here, we note that the in-plane weak FM moment is essential for the magnetic field control of the staggered moment axis. Indeed, the magnetization hysteresis curve obtained in $B \parallel [2\bar{1}\bar{1}0]$ at $T$ between 5 and 300 K reveals that a weak FM moment ($6$-$8$ m$\mu_{\rm B}$/Mn) changes its direction with almost the same coercivity as observed in the Hall effect [Fig. \ref{fig3}(a)]. While the in-plane $M$ is almost isotropic, exhibiting clear hysteresis, $M$ for $B \parallel [0001]$ mainly shows a linear $B$ dependence except a very tiny FM component of approximately $0.3$ m$\mu_{\rm B}$/Mn [Fig. \ref{fig3}(b)]. The in-plane weak ferromagnetism appears below $T_{\rm N} = 380$ K as can be seen in the $T$ dependence of the susceptibility $M/B$ for $B \parallel [2\bar{1}\bar{1}0]$ [Fig. \ref{fig3}(a) inset].

The Hall resistivity is conventionally described as the sum of the normal and anomalous Hall effects, which are proportional to $B$ and $M$, respectively. However, to characterize the spontaneous Hall effect seen in the noncollinear antiferromagnet Mn$_3$Sn \cite{Mn3Sn}, we find that the additional term $\rho_{\rm H}^{\rm AF}$ is necessary, and its Hall resistivity can be written as, 
\begin{eqnarray}
	\rho_{\rm H} = R_0 B + R_{\rm s} \mu_0 M + \rho_{\rm H}^{\rm AF},
\end{eqnarray} 
where $R_0$ and $R_{\rm s}$ are the normal and anomalous Hall coefficients, respectively. Here, we examine if the same Eq. (1) may describe the AHE in Mn$_3$Ge. The large zero-field component indicates that the AHE should dominate the Hall effect. To further confirm this, we estimate the normal Hall effect (NHE) using the field dependence of $\rho_{\rm H}$ at 400 K in the paramagnetic regime, where the in-plane and out-of-plane $\rho_{\rm H}(B)$ both linearly increase with $B$ with nearly the same slope (see Appendix B). The slope yields $R_{\rm H} = d\rho_{\rm H}/dB \sim 0.015 \ \mu \Omega$cm/T, which provides the upper limit of the estimate of $R_0$ and thus indicates that the NHE contribution is negligibly small and the AHE dominates $\rho_{\rm H}$ (see Appendix B). 

Next, to check the magnetization dependence of the AHE, we plot $\rho_{\rm H}$ vs $M$, taking the magnetic field as an implicit parameter [Fig. \ref{fig2}(d)]. For the $z$-axis component, $\rho_{\rm H}$ linearly increases with $M$ and, thus, $\rho_{\rm H}^{\rm AF} = 0$. For the $x$-$y$ plane component, $\rho_{\rm H}$ in a high-field regime also increases linearly with $M$ with a positive slope, $R_{\rm s} = d\rho_{\rm H}/dM$. However, in the low-field regime where $M(H)$ shows a hysteresis with a spontaneous component, the Hall resistivity also exhibits a hysteresis loop as a function of $M$. This is the same behavior as seen in Mn$_3$Sn \cite{Mn3Sn} and indicates that $\rho_{\rm H}$ has an additional spontaneous term $\rho_{\rm H}^{\rm AF} $ as described in Eq. (1). Notably, the magnetization in these two field regions has qualitatively different field response. The magnetization in the low-field regime corresponds to the weak ferromagnetism and exhibits hysteresis, while the high-field region with the small slope has the linear in-field increase, which most likely comes from the field-induced canting of the AF sublattices [Figs. \ref{fig3}(a) and \ref{fig3}(b)].

By using $R_0$ and the high-field $M$ slope, $R_{\rm s} = d\rho_{\rm H}/dM$, estimated above, we obtain $\rho_{\rm H}^{\rm AF} = \rho_{\rm H} - R_0 B - R_{\rm s} \mu_0 M$ as a function of both $M$ and $B$ [Figs. \ref{fig2}(e) and \ref{fig2}(f)]. Unlike the conventional AHE, $\rho_{\rm H}^{\rm AF}$ is not linearly dependent on $M$ or $B$. Given that the neutron-diffraction measurements and theoretical analysis show that the staggered moments of the chiral noncollinear spin structure freely rotate following the in-plane field \cite{Nagamiya1982,Tomiyoshi1983}, the large jump of $\rho_{\rm H}^{\rm AF}$ with a sign change in $M$ should come from the switching of the staggered moment direction. 

Normally, the AHE for a relatively resistive conductor is known to be proportional to the resistivity squared, $\rho^{2}$ \cite{nagaosa2010}. Thus, here we introduce the normalized parameter $S_{\rm H} = \mu_{0}R_{\rm s}/\rho^{2} = - \sigma_{\rm H}/M$. For FM conductors, $S_{\rm H}$ is independent of field and takes a value of the order of $0.01$-$0.1$ V$^{-1}$ \cite{nagaosa2010,Mn3Sn}. 
In high magnetic fields, $S_{\rm H} $ of Mn$_3$Ge indeed takes a constant value of approximately $+0.04~\mathrm{V^{-1}}$ (300 K), $-0.3  ~\mathrm{V^{-1}}$ (5 K) similar to ferromagnets \cite{Mn3Sn}. However, for the zero-field spontaneous component, we find strikingly large values $S_{\rm H}^0 = \rho_{\rm H}(B=0)/[\rho^2(B=0) M(B=0)] =  \mu_{0}R_{\rm s}^{\rm AF}/\rho^{2} \sim -1~\mathrm{V^{-1}}$ (300 K), $-8~\mathrm{V^{-1}}$ (5 K) for $B \parallel [01\bar{1}0]$. The extremely large value indicates that a distinct type of mechanism works here for the spontaneous Hall effect.

The temperature evolution of the spontaneous component of the AHE is examined by measuring the zero-field Hall resistivity $\rho_{\rm H}(B=0)$ and longitudinal resistivity $\rho(B=0)$ on heating. They are concomitantly measured in the FC condition, namely, after cooling the sample under a magnetic field $B_{\rm FC} = 7$ T from 350 K down to 5 K and consecutively setting $B \rightarrow 0$ at 5 K (see Appendix A). Figure \ref{fig4}(a) shows the $T$ dependence of the zero-field Hall conductivity $\sigma_{ji}(B=0) \approx -\rho_{ji}(B=0)/(\rho_{jj}(B=0)\rho_{ii}(B=0)$). The in-plane field components $|\sigma_{zy}|$ obtained after the Hall resistivity measurements in the FC condition in $B_{\rm FC} \parallel [2\bar{1}\bar{1}0]$ with $I\parallel [01\bar{1}0]$ and $|\sigma_{xz}|$  for $B_{\rm FC} \parallel [01\bar{1}0]$ and $I\parallel[0001]$ show nearly isotropic, large values reaching $310$ and $380~\Omega^{-1}$ cm$^{-1}$ at $T < 10$ K, respectively. Both $|\sigma_{zy}|$ and $|\sigma_{xz}|$ retain almost the same values up to approximately $50$ K where they start decreasing on heating. At 300 K, they remain nearly isotropic with $|\sigma_{zy}| = 40~\Omega^{-1}$ cm$^{-1}$ and $|\sigma_{xz}| = 55~\Omega^{-1}$ cm$^{-1}$ (see Appendix F), and finally vanish at $T_{\rm N} = 380$ K. In contrast, $|\sigma_{yx}|$ for $B_{\rm FC} \parallel [0001]$ and $I\parallel[01\bar{1}0]$ is less than 4 $\Omega^{-1}$ cm$^{-1}$ and remains much smaller than $|\sigma_{zy}|$ and $|\sigma_{xz}|$ at all $T \le T_{\rm N}$. 

Similarly, the longitudinal resistivity as a function of $T$ exhibits anisotropic behaviors [Fig. \ref{fig4}(a) inset]; the in-plane components with $I \parallel [2\bar{1}\bar{1}0]$ and $I\parallel[01\bar{1}0]$ overlap on top, peaking at 200 K and having relatively large residual resistivity $\rho(0) \sim 130~\mu\Omega$cm, while the out-of-plane component has a broad maximum at 300 K and shows a more conductive behavior with $\rho(0) \sim 50~\mu\Omega$cm. 
To estimate $S_{\rm H}^0 = -\sigma_{\rm H}(B = 0)/M(B = 0)$, we also measure $M(B = 0)$ in zero field after the same FC procedure using the same sample as those used for the Hall effect measurements. The in-plane field components of $S_{\rm H}^0$ are also found nearly isotropic [Fig. \ref{fig4}(b)], reaching a large value $> 5$ V$^{-1}$ at 5 K, 2 orders of magnitude larger than the values known for the conventional AHE \cite{nagaosa2010,Mn3Sn}.

The observed giant spontaneous Hall effect in an antiferromagnet is striking and indicates an unusual mechanism of the AHE. One can discuss the possible AHE based on a symmetry argument. The inverse chiral triangular spin structure reduces the symmetry of the lattice from the hexagonal to orthorhombic and, thus, may induce not only the weak ferromagnetism but the AHE in the $x$-$y$ plane. A numerical calculation using a different spin structure from the experimentally observed one indicates that the AHE can be large for Mn$_3$Ge \cite{Kubler2014}. The AHE is given by the Brillouin zone integration of the Berry curvature \cite{Niu2010}, and the significant contribution is found from the band-crossing points called Weyl points \cite{Wan2011,Burkov2011}. The large size of the observed anomalous Hall conductivity for in-plane field reaching approximately $310$-$380$ $\Omega^{-1}$cm$^{-1}$ under zero field has a similar magnitude as the theory, but the theory finds much more anisotropic AHE, as summarized in Table \ref{tab:comparison} in Appendix H \cite{Kubler2014}. The disagreement should come from the fact that the calculation in Ref. \cite{Kubler2014} was made using spin structures different from what is observed in experiment \cite{Nagamiya1982,Tomiyoshi1982,Tomiyoshi1983}. 

\begin{figure}[t]
	\begin{center}
		\includegraphics[width=6.9cm]{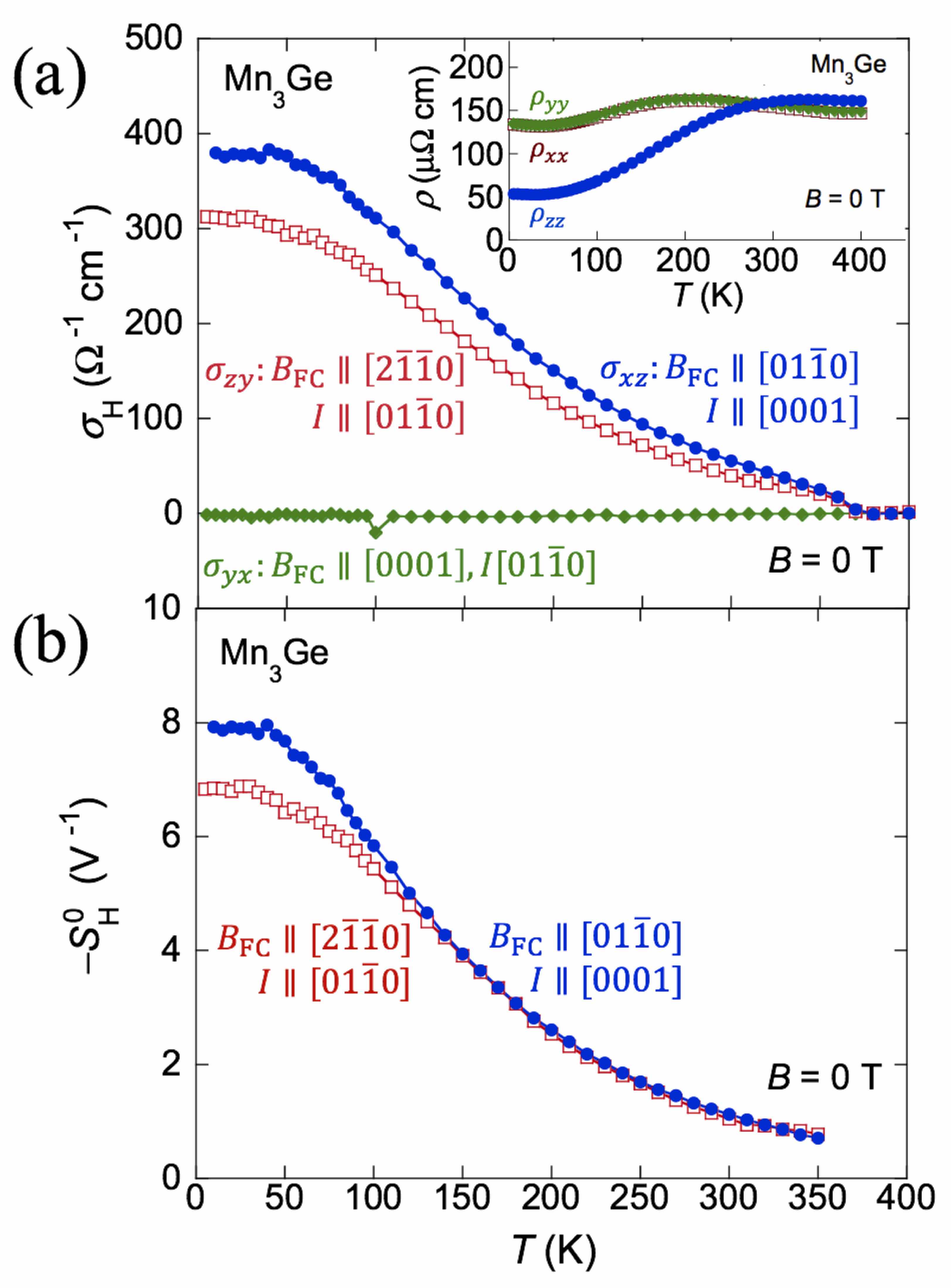}
		\caption{Temperature dependence of the anomalous Hall effect under zero field. All the data are obtained at zero field after the field-cooling (FC) procedures made in the magnetic field $B_{\rm FC}$. Directions of the field $B_{\rm FC}$ and electric current $I$ used for the Hall resistivity measurements are shown in each figure. (a)  Temperature dependence of the anomalous Hall conductivity $\sigma_{\rm H} (B = 0)$. The inset shows the temperature dependence of the longitudinal resistivity under zero field obtained after the same FC procedures. The same symbol and color as in the main panel are used for each FC configuration. $\rho_{xx}(T)$ (square) measured under zero field is also shown. (b) Temperature dependence of $S_{\rm H}^{0} = -\sigma_{\rm H}(B=0)/M(B=0)$ obtained at zero field after the FC procedures.}
		\label{fig4} 
	\end{center}
\end{figure}

Theoretically, the anomalous Hall conductivity of a 3D system can reach a value as large as the one known for a layered 3D QHE, which has been proposed to appear in the systems called Chern insulators. Notably, the zero-field AHE observed in Mn$_3$Ge reaches nearly half of $\sigma_{\rm H} =\frac{e^2}{2\pi h}|{\bf G}| \sim 800\ \Omega^{-1}$ cm$^{-1}$, a value expected for a 3D QHE with Chern number of unity where the pair of Weyl points are separated by the reciprocal lattice vector {\bf G} \cite{Yang2011,Turner2012}. The fact that the sizes of $|\sigma_{zy}|$ and $|\sigma_{xz}|$ are comparable suggests that the separation between the Weyl points must be similar to each other for the cases of $B_{\rm FC} \parallel [2\bar{1}\bar{1}0]$ and $[01\bar{1}0]$. 
On the other hand, the origin of the much smaller  $|\sigma_{yx}|$ than the in-plane field components $|\sigma_{zy}|$ and $|\sigma_{xz}|$ can be very small spin canting toward the $z$ axis and is a subject for future investigation.

\section*{III. Conclusion}
The large AHE observed in Mn$_3$Ge at room temperature may be significantly useful for various applications. In the field of spintronics, intensive studies have been made to find an antiferromagnet that serves as the active material for next-generation memory devices \cite{MacDonald2006, Jungwirth2010,MacDonald2011,Park2011,Marti2014,Gomonay2014}. In contrast with ferromagnets that have been mainly used for spintronics \cite{Fert2007}, antiferromagnets have robust stability against magnetic field perturbation and produce vanishingly small stray fields, thus, allowing high-density memory integration. The observed giant AHE in the chiral antiferromagnet Mn$_3$Ge with a very small magnetization indicates that the material has a large fictitious field (equivalent to $> 200$ T) in the momentum space without producing almost any perturbing stray fields. The fact that the large fictitious field may be readily controlled by the application of a low external field indicates that the antiferromagnet will be useful, for example, to develop various switching and memory devices.

\section*{acknowledgments}
\begin{acknowledgments}
We thank Tomoya Higo, Muhammad Ikhlas, Hidetoshi Fukuyama, and Ryotaro Arita for useful discussions.
This work is partially supported by PRESTO and CREST, JST, Grants-in-Aid for Scientific Research (Grants No. 25707030 and No.16H02209), by Grants-in-Aids
for Scientific Research on Innovative Areas (Grants No. 15H05882 and No. 15H05883), and the Program for Advancing Strategic International Networks to Accelerate the Circulation of Talented Researchers (Grant No. R2604) from JSPS. 
\end{acknowledgments}

\begin{table*}[t]
	\caption{\label{tab:table1} Crystal structure parameters refined by Rietveld analysis for $\epsilon$-Mn$_{3.05}$Ge$_{0.95}$ with $P6_3/mmc$ structure at 300 K. The lattice parameters and the atomic positions of the Mn site are determined by the analysis, which is made using the x-ray diffraction spectra with CuK$\alpha$ radiation ($\lambda = 1.5418$ \AA). The final $R$ indicators are $R_{\rm WP}$=1.83, $R_{\rm e}$=1.23, and $S$=1.53 \cite{Izumi2007}.}
\begin{ruledtabular}
	\begin{tabular}{lrllll}
		\multicolumn{2}{l}{Mn$_{3+x}$Ge$_{1-x}$ ($x=0.05$)} &\multicolumn{4}{c}{$V=106.51(4)$ \AA $^3$} \\
		\hline
		\multicolumn{1}{l}{lattice parameters (Spacegroup : $P6_3/mmc$)} &\multicolumn{1}{c}{} & \multicolumn{1}{c}{$a=5.338(1)$ \AA} & \multicolumn{1}{c}{$b=5.338(1)$ \AA  } & \multicolumn{1}{c}{ $c=4.3148(3)$ \AA  }&\multicolumn{1}{c}{} \\
		\hline
		\multicolumn{1}{l}{ Atom                    } &\multicolumn{1}{c}{Wyckoff position} & $x$ & $y$ &  $z$& Occupancy\\
		\multicolumn{1}{l}{ Mn                   } &\multicolumn{1}{c}{6h} &  0.833(1) &  0.666(2) &   1/4& 1\\
		\multicolumn{1}{l}{ Ge/Mn               } &\multicolumn{1}{c}{2c} &  1/3 &  2/3 &   1/4& (0.95/0.05)\\
	\end{tabular}
	\end{ruledtabular}
	\label{tab:Ret}
\end{table*}

\it{Note added.}\rm---After the completion of our work, we became aware of a similar work by Nayak {\it et al}. \cite{Nayak2015}. The strong anisotropy for the in-plane field configuration in the Hall conductivity in Ref. \cite{Nayak2015} is inconsistent with our results. This most likely comes from the fact that we take account of the observed anisotropy of the longitudinal resistivity in the analysis of the Hall conductivity, as detailed in Appendix E.

\begin{figure*}[t]
\begin{minipage}[c]{0.77\textwidth}
		\includegraphics[keepaspectratio, width=\textwidth]{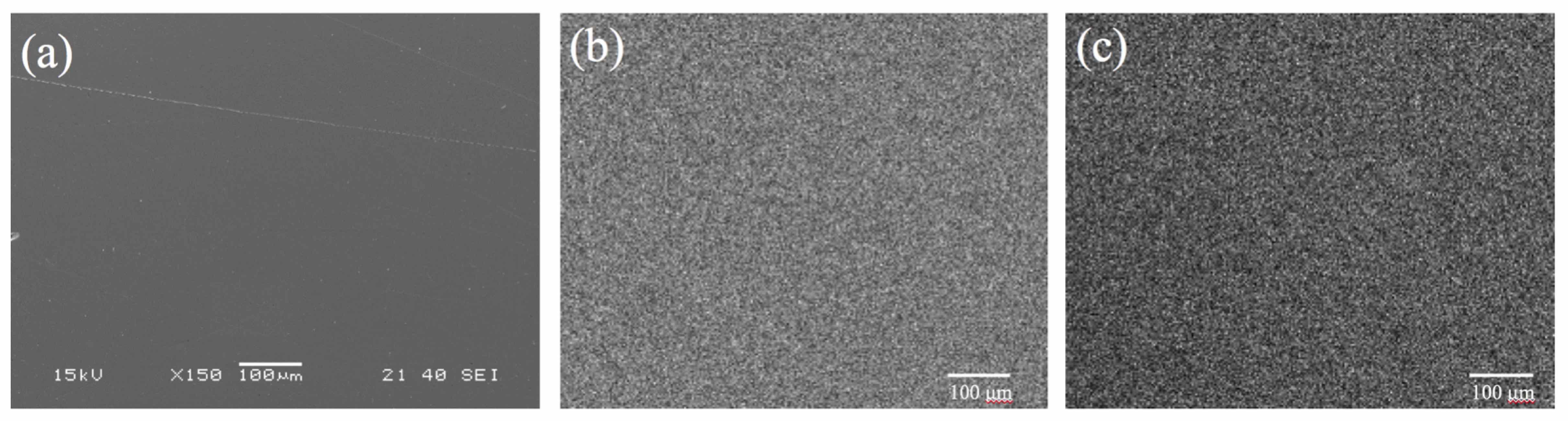}
		\end{minipage}\hfill
  \begin{minipage}[c]{0.2\textwidth}
		\caption{Room-temperature SEM EDX analyses for a Mn$_3$Ge single crystal. (a) SEM image of a polished surface and associated (b) Mn and (c) Ge EDX mapping are shown.  The maps are obtained with accelerating voltage of 15 kV.}\label{edx}
	\end{minipage}
\end{figure*}

\section*{Appendix A: Experiment}

Polycrystalline samples are prepared by arc melting the mixtures of manganese and germanium in a purified argon atmosphere. Excess manganese (12 mol \%) over the stoichiometric amount is added to compensate the loss during the arc melting and the crystal growth. The obtained polycrystalline materials are used for crystal growth by the Czochralski method using a commercial tetra-arc furnace (TAC-5100, GES). Subsequently, the sample is annealed for three days at 860 $^\circ\mathrm{C}$ and quenched in water in order to remove the low-temperature phase, which has the tetragonal Al$_3$Ti-type structure. Our SEM EDX (scanning electron microscopy with energy-dispersive x-ray spectroscopy) analysis for single crystals indicates that Mn$_3$Ge is the bulk phase, and we find that the composition of the single crystals is Mn$_{3.05}$Ge$_{0.95}$ (Mn$_{3.22}$Ge).
Our single-crystal and powder x-ray measurements at 300 K confirm the majority of the hexagonal $\varepsilon$ phase ($P6_3/mmc$) of Mn$_3$Ge with a small inclusion of the tetragonal phase whose volume fraction is less than 1\%. This is consistent with our observation of a ferromagnetic component of approximately $0.001~\mu_{\rm B}$/f.u. in the magnetization curve under $B \parallel c$ [Fig. \ref{fig3}(b) in the main text], taking account of the fact that the tetragonal phase is ferrimagnetic and has the net magnetization of $\ge 0.4~\mu_{\rm B}$/f.u. at room temperature \cite{Kurt2012}.  Rietveld analysis is made for the hexagonal phase of Mn$_3$Ge and the associated results shown in Table \ref{tab:Ret} agree with those in the literature \cite{Niida1993}. Figure \ref{edx} shows the SEM EDX mapping of a polished surface of a Mn$_3$Ge single crystal. The EDX mapping images for Mn and Ge show that Mn and Ge are homogeneously mixed. In this paper, we mainly report the results on the crystal whose composition is Mn$_{3.05}$Ge$_{0.95}$ (Mn$_{3.22}$Ge), and we refer to the crystal as ``Mn$_3$Ge" for clarity throughout the paper. On the other hand, in order to investigate the composition dependence of the Hall resistivity [Appendix D, Fig. \ref{CompDep}], we also grow a single crystal whose composition is Mn$_{3.07}$Ge$_{0.93}$ (Mn$_{3.32}$Ge).

\begin{figure*}[t]
\begin{minipage}[c]{0.77\textwidth}
		\includegraphics[keepaspectratio, width=14cm]{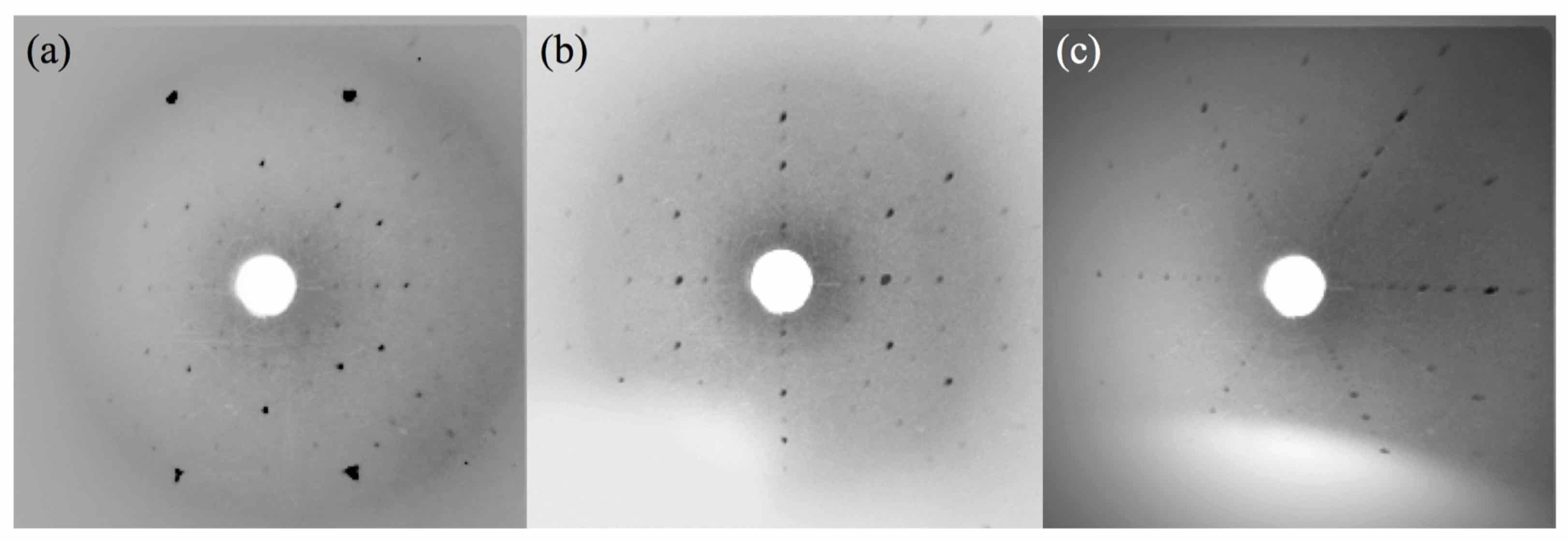}
		
		\end{minipage}\hfill
  \begin{minipage}[c]{0.2\textwidth}
		\caption{Laue images of a Mn$_3$Ge single crystal for  (a) $(2\bar{1}\bar{1}0)$, (b) $(01\bar{1}0)$, and (c) $(0001)$ orientations.}\label{laue}
\end{minipage}
\end{figure*}

We measure the resistivity and magnetization using annealed single crystals after making a bar-shaped sample through the alignment made by using a Laue diffractometer (Fig. \ref{laue}). We perform the magnetization measurements using a commercial superconducting quantum-interface-device magnetometer (MPMS, Quantum Design). We measure both longitudinal and Hall resistivities by a standard four-probe method using a commercial measurement system (PPMS, Quantum Design). In all the measurements, directions of the magnetic field, electric current, and Hall voltage are set perpendicular to each other.

\begin{figure*}[t]
\begin{minipage}[c]{0.77\textwidth}
		\includegraphics[keepaspectratio, height=6cm]{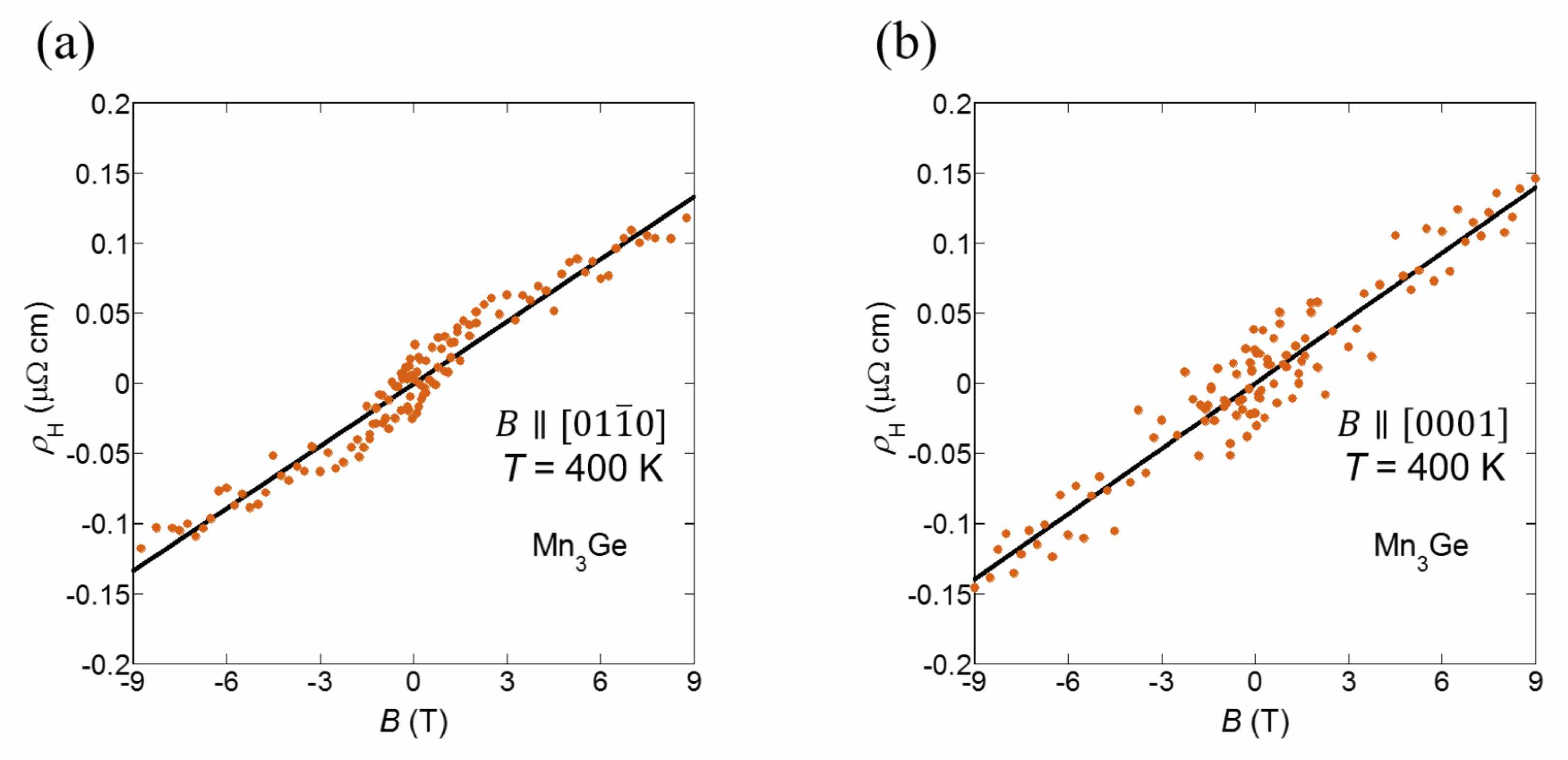}
				\end{minipage}\hfill
  \begin{minipage}[c]{0.2\textwidth}
		\caption{{Estimation of the carrier density based on the field dependence of the Hall resistivity. Hall resistivity $\rho_\mathrm{H}$ versus the magnetic field $B$ measured (a) in $B\parallel [01\bar{1}0]$ with $I\parallel [0001]$  and (b)  $B\parallel  [0001]$ with $I\parallel [01\bar{1}0]$  at 400 K. Black solid lines show the linear fit to estimate the carrier density $n$.}
		}\label{Hallres400}
\end{minipage}
\end{figure*}

We estimate the zero-field component of the anomalous Hall effect shown in Fig. \ref{fig4} in the main text by the following method. We cool down samples from 400 K down to 5 K under a field of $B_{\rm FC} =7$ T ($-7$ T), and, subsequently, at 5 K, we decrease the field $B$ down to $+0$ T ($-0$ T) without changing the sign of $B$. Then, we measure the Hall voltage $V_{\rm H}(B \rightarrow +0)$ ($V_{\rm H}(B \rightarrow -0)$) in zero field at various temperatures on heating after stabilizing temperature at each point. To remove the longitudinal resistance component induced by the misalignment of the Hall voltage contacts, we estimate the zero-field component of the Hall resistance as $R_{\rm H} (B = 0) = [V_{\rm H}(B \rightarrow +0) - V_{\rm H}(B \rightarrow -0)]/2I$. Here, $I$ is the electric current. Different samples are used for each field-cooling configuration shown in Fig. \ref{fig4} in the main text. We measure the longitudinal resistivity at zero field $\rho(B = 0)$ concomitantly in the same procedures as those used for the Hall resistivity measurements. In addition, the zero-field longitudinal resistivity is measured using neighboring parts cut from the same crystal, and all the results and their anisotropy are consistent with those in Fig. \ref{fig4}(a) inset within an error bar of 10\%. We also measure the zero-field remanent magnetization $M(B = 0)$ using the same field-cooling procedures, and the same samples as used in both longitudinal and Hall resistivity measurements.

\section*{Appendix B: Estimate of carrier concentration and fictitious field}

The field dependence of the Hall resistivity at 400 K $> T_{\rm N}$ was obtained after subtracting the longitudinal resistivity component. Figures \ref{Hallres400}(a) and \ref{Hallres400}(b), respectively, show the Hall resistivity $\rho_\mathrm{H}$ versus $B$ measured in $B\parallel [01\bar{1}0]$ and $[0001]$ obtained at 400 K, which we find almost the same as each other. Black solid lines indicate linear fits yielding the slope $R_{\rm H} = d\rho_{\rm H}/dB \sim 0.015 \ \mu \Omega$cm/T for both orientations. Given a field-induced AHE contribution, this value of $R_{\rm H}$ provides the upper limit of the estimate of the normal Hall coefficient $R_{\rm 0}$ and, thus, corresponds to the lower bound for the carrier concentration, namely, $n \sim 4 \times 10^{22}$ /Mn.
 The fictitious magnetic field corresponding to Berry curvature in $k$ space can be estimated using $B_{\rm f}=|\rho^{\rm AF}_{\rm H}/R_0|$, where $R_{\rm H}  \sim 0.015 \ \mu \Omega$cm/T is used as the upper limit of the normal Hall coefficient $R_0$. For example, since $\rho^{\rm AF}_{\rm H} \sim 3 \ \mu\Omega\mathrm{cm}$ at 5 K [Fig. \ref{fig2}(f) in the main text], the fictitious field $B_{\rm f}$ should be higher than approximately $200 \ \mathrm{T}$. 

\begin{figure}[h]
	\begin{center}
		\includegraphics[keepaspectratio, height=6.5cm]{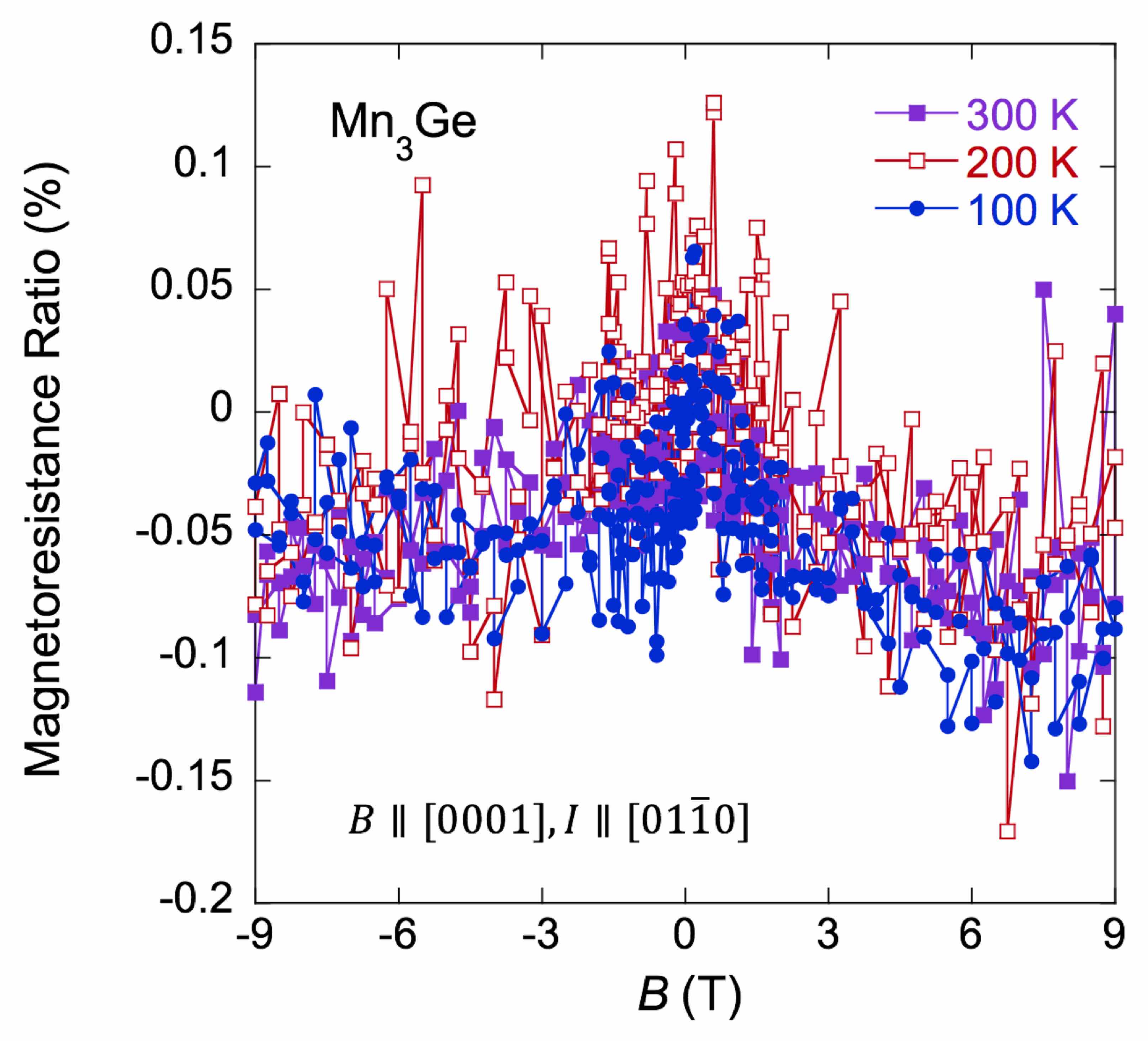}
		\caption{Field dependence of the longitudinal magnetoresistance ratio $
(\rho(B)-\rho(B = 0) )/\rho(B = 0)$ at various temperatures in the magnetic field $B\parallel[0001]$ with electric current $I\parallel [01\bar{1}0]$.}\label{magnetores}
\end{center}
\end{figure}

\section*{Appendix C: Field dependence of the longitudinal resistivity}

\begin{table*}[t]
	\begin{ruledtabular}
		\caption{Data used to estimate the anisotropic Hall conductivity $\sigma_{\rm H}\equiv \sigma_{ji} \approx -\rho_{ji}/\rho_{jj}\rho_{ii}$. We find that for all field directions, the magnetoresistance is much less than 1\% at all $T$ ranges between 5 and 400 K. Therefore, the field-dependent longitudinal resistivity is approximated into the same constant value as the zero-field one.}
		\begin{tabular}{lcl}   
			
			$\sigma_{\rm H}\equiv \sigma_{ji} \approx -\rho_{ji}/\rho_{jj}\rho_{ii}$ & Concomitantly measured $\rho_{ji}$ and $\rho_{ii} (\rho_{jj})$ & $\rho_{jj} (\rho_{ii})$ data used for $\sigma_\mathrm{H}$\\
			\hline
			$\sigma_{yx}$ in Fig. \ref{fig2}(c)	&$\rho_{yx}$ and $\rho_{yy}$& $\rho_{xx}$: 5 K data in Fig. \ref{fig4}(a) inset\\
			$\sigma_{xz}$ in Fig. \ref{fig2}(c)	&$\rho_{xz}$ and $\rho_{zz}$& $\rho_{xx}$: 5 K data in Fig. \ref{fig4}(a) inset\\ 
			$\sigma_{zy}$ in Fig. \ref{fig2}(c)	&$\rho_{zy}$ and $\rho_{yy}$ & $\rho_{zz}$: 5 K data in Fig. \ref{fig4}(a) inset\\
			$\sigma_{yx}$ in Fig. \ref{fig4}(a)	&$\rho_{yx}$ and $\rho_{yy}$& $\rho_{xx}$: $T$-dependent data in Fig. \ref{fig4}(a) inset\\
			$\sigma_{xz}$ in Fig. \ref{fig4}(a)	&$\rho_{xz}$ and $\rho_{zz}$& $\rho_{xx}$: $T$-dependent data in Fig. \ref{fig4}(a) inset\\
			$\sigma_{zy}$ in Fig. \ref{fig4}(a)	&$\rho_{zy}$ and $\rho_{yy}$& $\rho_{zz}$: $T$-dependent data in Fig. \ref{fig4}(a) inset\\
			$\sigma_{yx}$ in Fig. \ref{HallCond300}&$\rho_{yx}$ and $\rho_{yy}$& $\rho_{xx}$: 300 K data in Fig. \ref{fig4}(a) inset\\
			$\sigma_{xz}$ in Fig. \ref{HallCond300}&$\rho_{xz}$ and $\rho_{zz}$& $\rho_{xx}$: 300 K data in Fig. \ref{fig4}(a) inset\\ 
			$\sigma_{zy}$ in Fig. \ref{HallCond300}	&$\rho_{zy}$ and $\rho_{yy}$ & $\rho_{zz}$: 300 K data in Fig. \ref{fig4}(a) inset\\
		\end{tabular}
		\label{tab:HallCondOrgn}
	\end{ruledtabular}
\end{table*}
\begin{figure*}[t]
\begin{minipage}[c]{0.77\textwidth}
		\includegraphics[keepaspectratio, height=6.5cm]{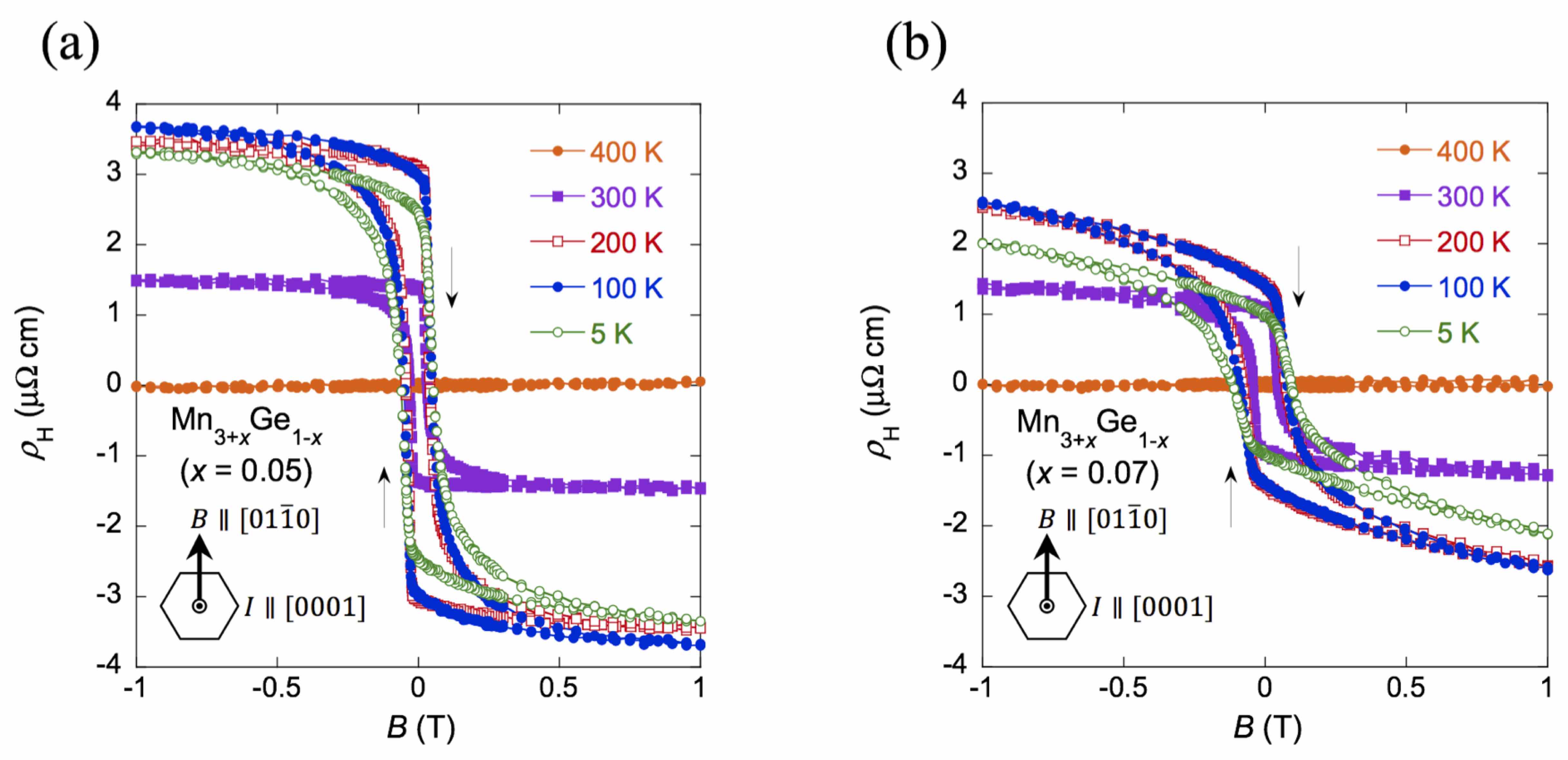}
		
				\end{minipage}\hfill
  \begin{minipage}[c]{0.2\textwidth}
		\caption{Field dependence of the Hall resistivity of (a) Mn$_{3.05}$Ge$_{0.95}$ (Mn$_{3.22}$Ge) and (b) Mn$_{3.07}$Ge$_{0.93}$ (Mn$_{3.32}$Ge) single crystals obtained at various temperatures in the magnetic field $B \parallel [01\bar{1}0]$ with the electric current $I \parallel [0001]$.}\label{CompDep}
	\end{minipage}
\end{figure*}

Over all temperature regions, a very small magnetoresistance is observed for all the in-plane and out-of-plane field directions. 
We find the magnetoresistance ratio $(\rho(B)-\rho(B = 0) )/\rho(B = 0)$ to be much less than 1\% and the associated resistivity is less than 10\% of the Hall resistivity change. For example, in Fig. \ref{magnetores}, we show the magnetoresistance ratio at various temperatures in the magnetic field $B\parallel[0001]$ with $I\parallel [01\bar{1}0]$. 

\section*{Appendix D: Composition dependence of the Hall resistivity}
We find a large difference in the Hall resistivity $\rho_{\rm H}$ at $|B|<1$ T between Mn$_{3.05}$Ge$_{0.95}$ and Mn$_{3.07}$Ge$_{0.93}$, as shown in Fig. \ref{CompDep}. The difference in $\rho_{\rm H}$ becomes larger with lowering temperature. For example, at zero field, the Hall resistivity $\rho_{\rm H}$ at 5 K in Mn$_{3.05}$Ge$_{0.95}$ is twice larger than in Mn$_{3.07}$Ge$_{0.93}$. On the other hand, a similar magnitude is seen in $\rho_{\rm H}$ for the results obtained above $T=300$ K. The coercivity increases with $x$ from approximately $200$ Oe ($x=0.05$) to approximately $500$ Oe ($x=0.07$), indicating that the amount of the lattice defects and disorder increases with more excess of Mn.

\section*{Appendix E: Estimate of anisotropic Hall conductivity}
The Hall conductivity is estimated as $\sigma_{\rm H}\equiv \sigma_{ji} \approx -\rho_{ji}/\rho_{jj}\rho_{ii}$ taking account of the observed anisotropy of the longitudinal resistivity, where $(i,j)=(x,y), (y,z),$ or $(z,x)$. 
For this analysis, two sets of the transport results are necessary (see Table \ref{tab:HallCondOrgn}). One is the Hall resistivity $\rho_{ji}$ and the longitudinal resistivity $\rho_{ii}$ ($\rho_{jj}$), both of which are concomitantly measured as described in Appendix A. The other is the longitudinal resistivity $\rho_{jj}$ ($\rho_{ii}$) for the vertical direction to $\rho_{ii}$ ($\rho_{jj}$). To estimate the intrinsic anisotropy, both $\rho_{ii}$ and $\rho_{jj}$ are measured using the same sample or neighboring parts cut from the same crystal, as we describe above.
Because of the anisotropy in the longitudinal resistivity, $\sigma_{\rm H}$ can be overestimated or underestimated from the one using the above equation if we calculate the Hall conductivity as $\sigma_{\rm H}=-\rho_{\rm H}/\rho^2$. For example, in our measurements, $-\rho_{xz}/\rho_{zz}^2$ reaches $\simeq 950 \ \Omega^{-1}\ \mathrm{cm}^{-1}$ at $T < 50$ K, and this is more than twice a larger value than $\sigma_{xz} \approx -\rho_{xz}/\rho_{xx}\rho_{zz} \sim 380~\Omega^{-1}\ \mathrm{cm}^{-1} $ estimated using anisotropic longitudinal resistivity in the same $T$ range.

\begin{table*}[t]
	\begin{ruledtabular}
		\caption{Comparison between our experimental work and theoretical calculation by K\"ubler and Felser \cite{Kubler2014}. The anomalous Hall conductivity ($\sigma_{yx}$, $\sigma_{zy}$, and $\sigma_{xz}$) under zero field is listed for various temperatures based on the results shown in Fig. \ref{fig4}(a) in the main text. The results of the theoretical calculations using the spin configurations in Figs. 2, 3(b), 5(c), and 7 of Ref. \cite{Kubler2014} are also listed, where the asterisk(*) indicates the case with no spin-orbit interaction. The definition of the $x$ and $y$ coordinates used in Ref. \cite{Kubler2014} differs from those used in this paper. Therefore, the Hall conductivity results of Ref. \cite{Kubler2014} are listed after the coordinate transformation is applied from their definition to our definition, i.e., $x\parallel[2\,\bar{1}\,\bar{1}\,0]$, $y\parallel[0\,1\,\bar{1}\,0]$, and $z\parallel[0\,0\,0\,1]$.}
		\begin{tabular}{clrrrrrr}   
			
			Material & Spin configuration & $T$ [K] & $\sigma_{yx} [\mathrm{1/(\Omega cm)}]$ & $\sigma_{zy} [\mathrm{1/(\Omega cm)}]$ & $\sigma_{xz} [\mathrm{1/(\Omega cm)}]$ \\
			\hline
			\multirow{4}{*}{\shortstack{$\varepsilon$-Mn$_3$Ge\\(This work)}} & &5  &$ -1$ &$310$ & $380$\\
			& &100   &$-3$ &$250$ &$310$ \\
			& &200   &$ -3$ &$120$ &$150$ \\
			& &300  &$-1$ &$40$ &$55$ \\ \hline
			\multirow{5}{*}{\shortstack{$\varepsilon$-Mn$_3$Ge\\(Theoretical work \cite{Kubler2014})
				}} &Fig. 2, Ref. \cite{Kubler2014}&   &$0$ &$0$ &$0$ \\
				&Fig. 3(b), Ref. \cite{Kubler2014}  &     &$0$ &$-379$ &$-667$ \\
				&Fig. 5(c), Ref. \cite{Kubler2014}  &     &$0$ &$-1$ &$-607$ \\
				&Fig. 7, Ref. \cite{Kubler2014}      &      &$-104$ &$965$ &$-231$ \\
				&Fig. 7, Ref. \cite{Kubler2014}$^*$ &   &$-85$ &$-4$ &$-6$ \\
			\end{tabular}
			\label{tab:comparison}
	\end{ruledtabular}
	\end{table*}

\section*{Appendix F: Anisotropy in field dependence of the Hall conductivity at 300 K}
\begin{figure}[h]
	\begin{center}
		\includegraphics[keepaspectratio, height=6.5cm]{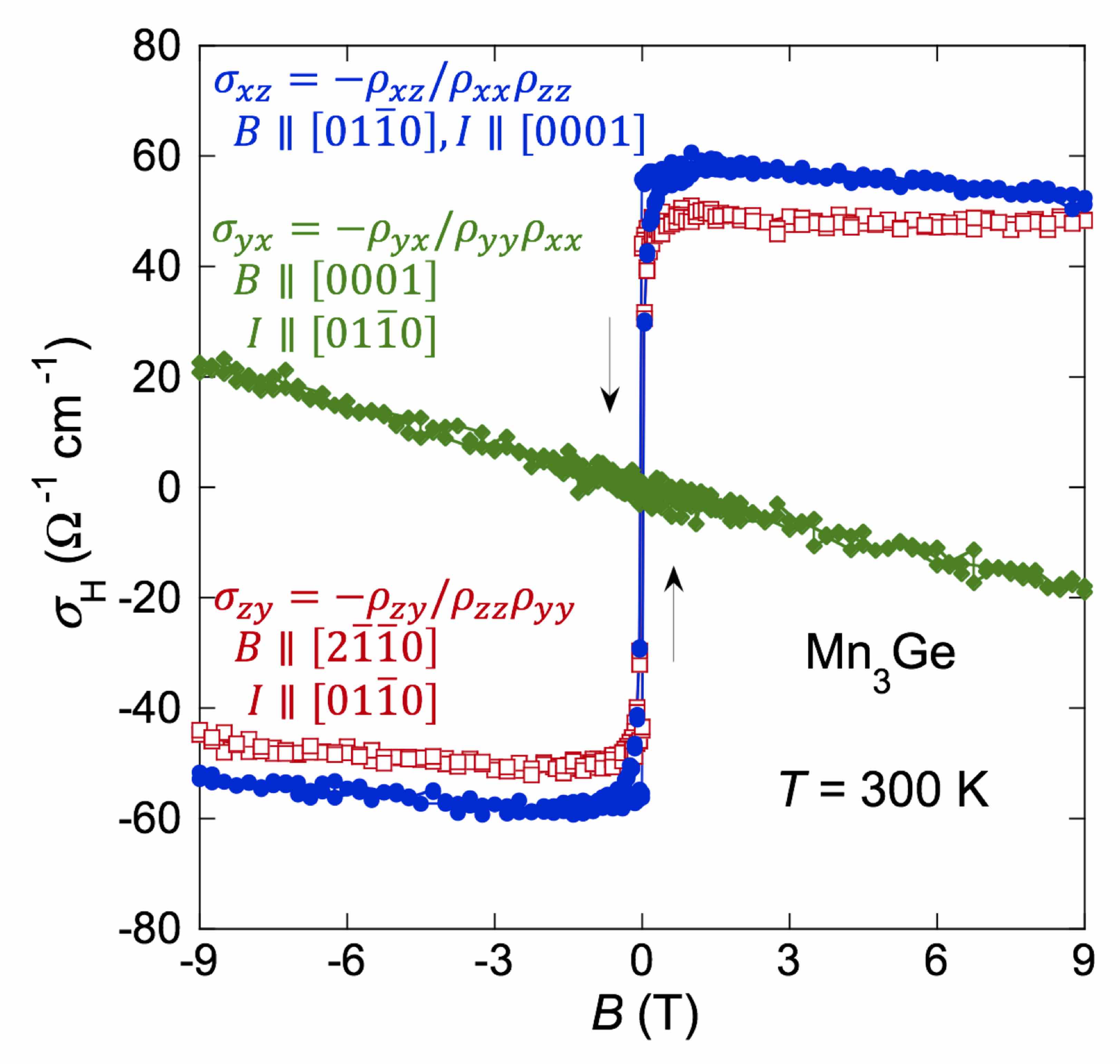}
		\caption{Anisotropic field dependence of the Hall conductivity obtained through a field cycle at 300 K. Direction of the magnetic field $B$ and electric current $I$ used for Hall resistivity measurements are shown.
		}\label{HallCond300}
	\end{center}
\end{figure}

The field dependence of the Hall conductivity $\sigma_\mathrm{\rm H}$ at 300 K for the field along the $x$-$y$ plane and the $z$ axis is shown in Fig. \ref{HallCond300}. For the in-plane field, $\sigma_\mathrm{\rm H}$ at 300 K is nearly isotropic, similar to the results at 5 K in Fig. \ref{fig2}(c) in the main text. 

\section*{Appendix G: Spin configurations in in-plane magnetic field}

\begin{figure}[h]
	\begin{center}
		\includegraphics[keepaspectratio, width=\columnwidth]{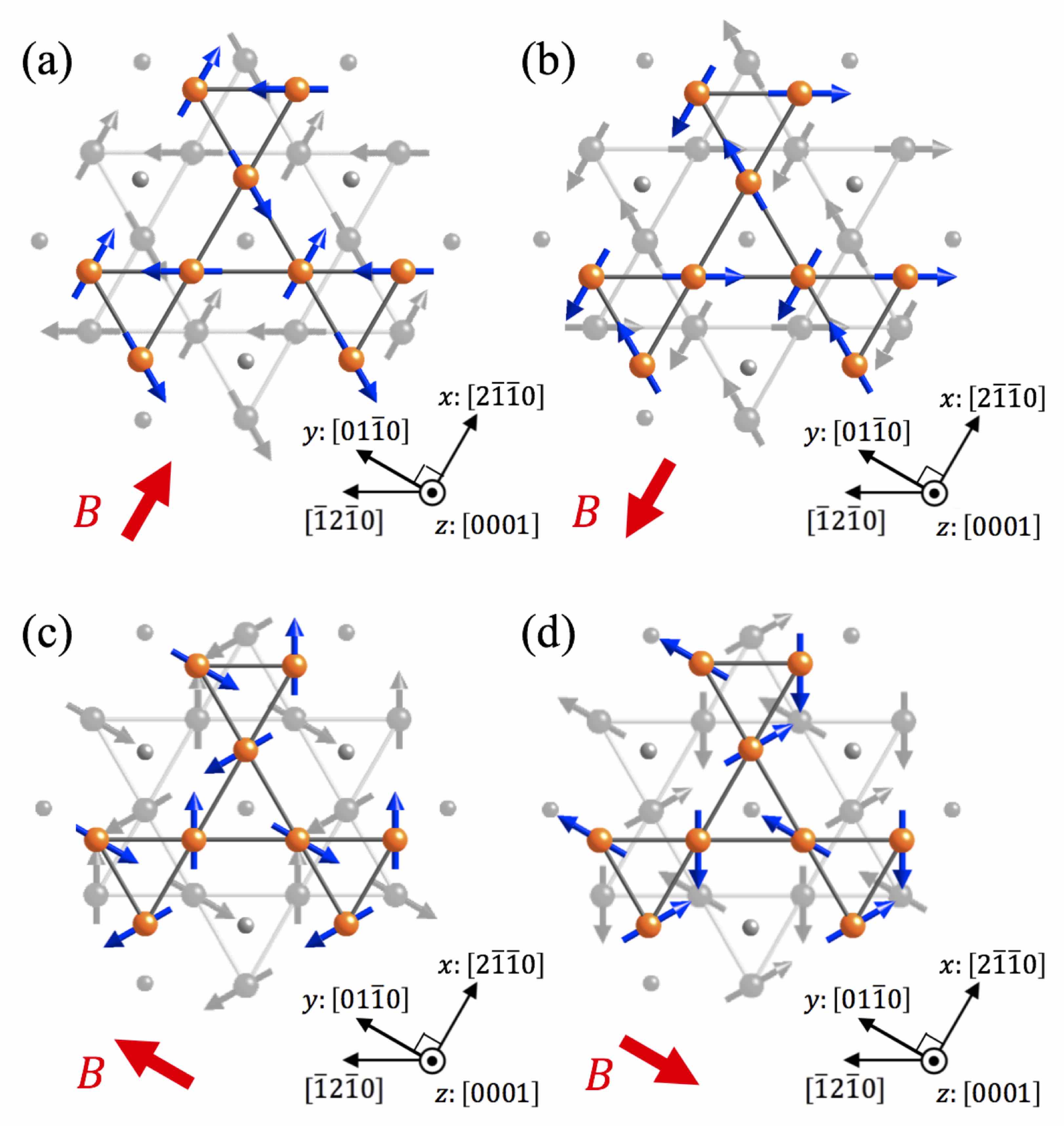}
		\caption{Spin configurations in Mn$_3$Ge at the $z=0$ plane (gray) and the $z=1/2$ plane (color) under the field applied  along (a) $[2\bar{1}\bar{1}0]$, (b) $[\bar{2}110]$, (c) $[01\bar{1}0]$, and (d) $[0\bar{1}10]$. The red arrows indicate the direction of the magnetic field $B$.
		}\label{spinflip}
	\end{center}
\end{figure}

When the magnetic field is applied along the in-plane directions $[2\bar{1}\bar{1}0]$ and $[01\bar{1}0]$, the change in the spin configuration occurs from the one in Fig. \ref{spinflip} (a) to the other in Fig. \ref{spinflip}(b), and from Fig. \ref{spinflip}(c) to Fig. \ref{spinflip}(d), respectively. The corresponding jumps in the Hall signal and magnetization are observed as a function of field as shown in Figs. \ref{fig2} and \ref{fig3} in the main text, respectively.

\section*{Appendix H: Comparison between theoretical calculations and experimental results}
In Table \ref{tab:comparison}, we compare our results of the anomalous Hall conductivity of Mn$_3$Ge with those calculated for spin configurations by K\"ubler and Felser \cite{Kubler2014} (Figs. 2, 3, 5, and 7). While the order of magnitude is similar, our results are different from their calculations in terms of the sign and anisotropy. It should be noted that spin configurations used in Ref. \cite{Kubler2014} are not consistent with the results obtained from the neutron-diffraction measurements \cite{Nagamiya1982,Tomiyoshi1983}.

\bibliography{Mn3Ge_PRApplied_for_arXiv_v2}
\end{document}